\documentclass{elsart3p}
\usepackage{epsfig}
\usepackage{amsmath}
\usepackage{amssymb}
\usepackage{amsfonts}
\usepackage{mathptmx}
\usepackage{eucal}
\usepackage{bm}

\DeclareMathOperator{\re}{\mathop{\mathrm{Re}}}

\newcommand{\Eq}[1]{Eq.~(\ref{#1})}
\newcommand{\Eqs}[1]{Eqs.~(\ref{#1})}

\begin{document}

\begin{frontmatter}

\title{Dissipative current in SIFS Josephson junctions}

\author[label1,label2]{A.~S.~Vasenko},
\ead{Andrey.Vasenko@grenoble.cnrs.fr}
\author[label3]{S.~Kawabata},
\author[label4]{A.~A.~Golubov},
\author[label5]{M.~Yu.~Kupriyanov},
\author[label1]{F.~W.~J.~Hekking}
\address[label1]{LPMMC, Universit\'{e} Joseph Fourier and CNRS, 25 Avenue des Martyrs, BP 166,
38042 Grenoble, France}
\address[label2]{Department of Physics, Moscow State University, Moscow 119992, Russia}
\address[label3]{Nanotechnology Research Institute (NRI), National Institute of Advanced
Industrial Science and Technology (AIST),
\\
Tsukuba, Ibaraki, 305-8568, Japan}
\address[label4]{Faculty of Science and Technology and MESA$^+$ Institute for Nanotechnology,
University of Twente, 7500 AE Enschede, The Netherlands}
\address[label5]{Nuclear Physics Institute, Moscow State University, Moscow, 119992, Russia}

\begin{abstract}
We investigate superconductor/insulator/ferromagnet/superconductor
(SIFS) tunnel Josephson junctions in the dirty limit, using  the
quasiclassical theory. We consider the case of a strong tunnel
barrier such that the left S layer and the right FS bilayer are
decoupled. We calculate quantitatively the density of states (DOS)
in the FS bilayer for arbitrary length of the ferromagnetic layer,
using a self-consistent numerical method. We compare these results
with a known analytical DOS approximation, which is valid when the
ferromagnetic layer is long enough. Finally we calculate
quantitatively the current-voltage characteristics of a SIFS
junction.
\end{abstract}

\end{frontmatter}

\section{Introduction}

It is well known that superconductivity and ferromagnetism are two
competing orders. The coexistence of singlet superconductivity and
ferromagnetism is basically impossible in the same compound but
may be easily achieved in artificially fabricated
superconductor/ferromagnet (S/F) hybrid structures. In this case,
the coexistence of the two orders is due to the proximity effect
\cite{RevB, RevG}. The main manifestation of the proximity effect
in S/F structures is the damped oscillatory behavior of
superconducting correlations in the F layer. Two characteristic
lengths of the decay and the oscillations are, respectively, $\xi
_{f1}$ and $\xi _{f2}$.

Therefore in S/F heterostructures there is a unique possibility
to study the properties of superconducting electrons under the
influence of the exchange field in the ferromagnet. Recent
progress in the preparation of the high quality S/F layered
structures permitted to experimentally observe many striking
phenomena that are quite interesting for applications such as a
non-monotonic dependence of their critical temperature and
oscillations of critical current in S/F/S Josephson junctions as a
function of the F layer thickness (see \cite{RevB} and references
therein). It is possible to fabricate Josephson $\pi$ junctions
with a $\pi$-phase difference in the ground state, which are good
candidates for elements in superconducting logic circuits
\cite{logic}.

SIFS junctions, i.e. S/F/S trilayers with one transparent
interface and one tunnel barrier between S and F layers, represent
an interesting case of $\pi$ junctions for applications where
active Josephson junctions are required. The SIFS structure offers
the freedom to tune the critical current density over a wide range
and at the same time to realize high values of the product of the
junction critical current $I_{c}$ and its normal state resistance
$R_{N}$ \cite{Weides}. In addition, Nb based tunnel junctions are
usually underdamped, which is desired for many applications. SIFS
junctions are also interesting from the fundamental point of view
since they provide a convenient model system for a comparative
study between $0$-$\pi$ transitions observed from the critical
current and from the density of states (DOS) \cite{Vasenko}.

The purpose of this work is to provide a quantitative model
describing the DOS in SIFS junctions and to calculate
current-voltage characteristics of a SIFS junction. The latter may
be used for estimation of the dissipation in SIFS-based qubit
systems.

\section{Model and basic equations}

\begin{figure}[tb]
\epsfxsize=7.5cm\epsffile{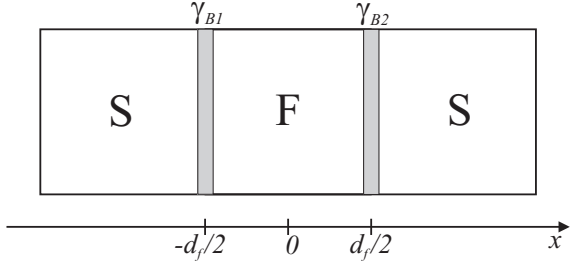} 
\caption{Geometry of the considered system. The thickness of the
ferromagnetic interlayer is $d_f$. The transparency of the left S/F
interface is characterized by the coefficient $\protect\gamma_{B1}$, and the
transparency of the right F/S interface is characterized by $\protect\gamma_{B2}$.}
\label{SIFS}
\end{figure}

The model of an S/F/S junction we are going to study is depicted
in Fig.~\ref{SIFS} and consists of a ferromagnetic layer of
thickness $d_{f}$ and two thick superconducting electrodes along
the $x$ direction. The left superconductor electrode is
voltage-biased. Left and right superconductor/ferromagnet
interfaces are characterized by the dimensionless parameters
$\gamma _{B1}$ and $\gamma _{B2}$, respectively, where $\gamma
_{B1,B2}=R_{B1,B2}\sigma _{n}/\xi _{n}$, $R_{B1,B2}$ are the
resistances of the left and right S/F interfaces, respectively,
$\sigma _{n}$ is the conductivity of the F layer, $\xi
_{n}=\sqrt{D_{f}/2\pi T_{c}}$, $D_{f}$ is the diffusion
coefficient in the ferromagnetic metal and $T_{c}$ is the critical
temperature of the superconductor (we assume $\hbar =k_{B}=1$). We
also assume that the S/F interfaces are not magnetically active.
We will consider the diffusive limit, in which the elastic
scattering length $\ell $ is much smaller than the characteristic
decay length $\xi _{f1}$.

In this paper we concentrate on the case of a SIFS tunnel
Josephson junction, when $\gamma _{B1}\gg 1$ (tunnel barrier) and
$\gamma _{B2}=0$ (fully transparent interface; however in our
numerical calculations we use finite but very small $\gamma _{B2}
\ll 1$). In this case the left S layer and the right FS bilayer in
Fig.~\ref{SIFS} are decoupled. Therefore we can calculate the
quasi-particle current through a SIFS junction using the standard
formula of the tunnel theory \cite{Werthammer},
\begin{equation}\label{I(V)}
I = \frac{1}{e R} \int_{-\infty}^{\infty}dE N_S(E - eV) N(E)
\left[ f(E - eV) - f(E) \right],
\end{equation}
where $N_S(E) = |E|\Theta(|E|-\Delta)/\sqrt{E^2 - \Delta^2}$ is
the BCS density of states [$\Theta(x)$ is the Heaviside step
function] and $N(E)$ is the density of states in the FS bilayer at
the free boundary of the ferromagnet ($x=-d_f/2$). Both $N_S(E)$
and $N(E)$ are normalized to their values in the normal state;
$f(E)=[1+\exp(E/T)]^{-1}$ is the Fermi function. To obtain $N(E)$
we should solve the Usadel equations in the ferromagnetic layer of
the FS bilayer.

Using the $\theta$-parameterizations of the normal and anomalous
Green functions, $G=\cos \theta $, $F=\sin \theta$, we can write
the Usadel equations in the F layer as \cite{Usadel, Demler}
\begin{equation}\label{Usadel}
\frac{D_{f}}{2} \frac{\partial ^{2}\theta _{f\uparrow (\downarrow
)}}{\partial x^{2}} = \left( \omega \pm ih + \frac{\cos
\theta_{f\uparrow (\downarrow )}}{\tau _{m}}\right) \sin
\theta_{f\uparrow (\downarrow )},
\end{equation}
where a positive (negative) sign in front of $h$ corresponds to
the spin up state $\uparrow $ (spin down state $\downarrow $),
$\omega =2 \pi T(n + \frac{1}{2})$ are the Matsubara frequencies,
$h$ is the exchange field in the ferromagnet, and the parameter
$\tau_m$ is the spin-flip scattering time. We consider a
ferromagnet with strong uniaxial anisotropy, in which case the
magnetic scattering does not couple the spin up and spin down
electron populations.

In the S layer the Usadel equations take the form (where we omit
subscripts `$\uparrow (\downarrow)$' because equations in
superconductor look identically for spin up and spin down electron
states),
\begin{equation}\label{Usadel_S}
\frac{D_s}{2} \frac{\partial^2 \theta_s}{\partial x^2} = \omega
\sin \theta_s - \Delta(x) \cos \theta_s.
\end{equation}
They should be completed with the self-consistency equation,
\begin{equation}
\Delta (x)\ln \frac{T_c}{T} = \pi T \sum\limits_{\omega > 0}
\left( \frac{2\Delta (x)}{\omega}-\sin \theta _{s \uparrow} - \sin
\theta _{s \downarrow} \right). \label{Delta}
\end{equation}
Here $D_s$ is the diffusion coefficient in the superconductor and $\Delta(x)$ is the superconducting pair potential.

Since we calculate the DOS at the free boundary of the ferromagnet in the FS bilayer,
we need to set to zero the $\theta_f$ derivative at the left S/F interface,
\begin{equation}\label{leftBK}
\left(\partial \theta_f/\partial x \right)_{-d_f/2} = 0.
\end{equation}
At the right F/S interface the boundary conditions are given by the relations \cite{KL},
\begin{subequations}
\label{KL}
\begin{align}
\xi_n\gamma\left( \partial \theta_f/\partial x \right)_{d_f/2} &=
\xi_s \left( \partial \theta_s/\partial x \right)_{d_f/2},
\label{KL1} \\
\xi_n \gamma_{B2} \left( \partial \theta_f/\partial x
\right)_{d_f/2} &= \sin\left( \theta_s - \theta_f \right)_{d_f/2},
\label{KL_DOS}
\end{align}
\end{subequations}
where $\gamma = \xi_s\sigma_n/\xi_n\sigma_s$, $\sigma_s$ is the
conductivity of the S layer and $\xi_s = \sqrt{D_s/2\pi T_c}$. The
parameter $\gamma$ determines the strength of suppression of
superconductivity in the right S lead near the interface compared
to the bulk: no suppression occurs for $\gamma=0$, while strong
suppression takes place for $\gamma \gg 1$. In our numerical
calculations we will assume small $\gamma \ll 1$.

To complete the boundary problem we also set a boundary condition at $x = \infty$,
\begin{equation}\label{gran1}
\theta_s(\infty) = \arctan\frac{|\Delta|}{\omega}.
\end{equation}

\section{Density of states in the FS bylayer}

\begin{figure}[tb]
\epsfxsize=7.5cm\epsffile{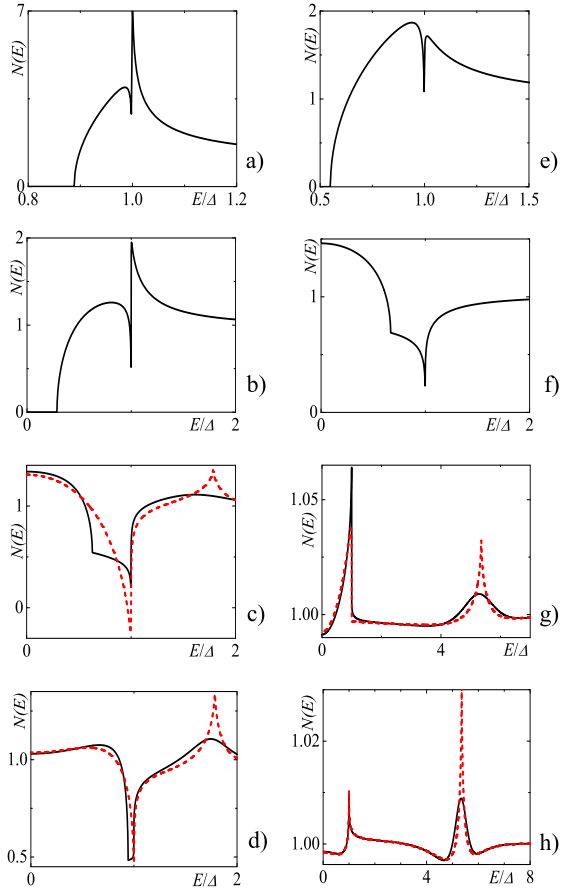} 
\caption{(Color online) DOS on the free boundary of the F layer in the FS bilayer calculated numerically in the
absence of spin-flip scattering. Plots (a)-(d) are calculated for $h = \pi T_c$ and (e)-(h) for $h = 3 \pi T_c$.
The temperature $T = 0.1 T_c$. Parameters of the F/S interface are $\gamma = \gamma_{B2} = 0.01$ (a),(e):
$d_f/\xi_n=0.5$; (b),(f): $d_f/\xi_n=1$; (c),(g): $d_f/\xi_n=2$; (d),(h): $d_f/\xi_n=3$. The approximate
analytical solution \cite{Vasenko} is shown by dashed red lines.}
\label{DOSa0}
\end{figure}
\begin{figure}[tb]
\epsfxsize=7.5cm\epsffile{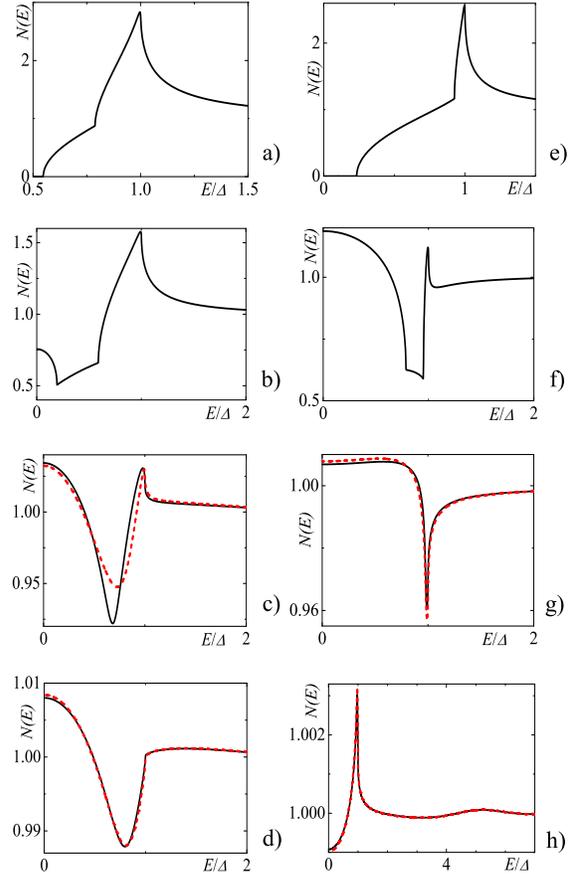} 
\caption{(Color online) DOS on the free boundary of the F layer in the FS bilayer calculated numerically for $\tau_m = 1/\pi T_c$.
Plots (a)-(d) are calculated for $h = \pi T_c$ and (e)-(h) for $h = 3 \pi T_c$.
The temperature $T = 0.1 T_c$. Parameters of the F/S interface are $\gamma = \gamma_{B2} = 0.01$ (a),(e):
$d_f/\xi_n=0.5$; (b),(f): $d_f/\xi_n=1$; (c),(g): $d_f/\xi_n=2$; (d),(h): $d_f/\xi_n=3$. The approximate
analytical solution \cite{Vasenko} is shown by dashed red lines.}
\label{DOSa1}
\end{figure}

To calculate the DOS $N(E)$ on the free F layer boundary
($x=-d_f/2$) we use the self-consistent two-step iterative
procedure \cite{GK1}. In the first step we calculate the pair
potential's coordinate dependence $\Delta(x)$ using the
self-consistency equation in the S layer, \Eq{Delta}. Then, by
proceeding to the analytical continuation in \Eqs{Usadel},
\eqref{Usadel_S}, \eqref{leftBK}, \eqref{KL}, \eqref{gran1} of the
quasi-particle energy $i\omega \rightarrow E + i0$ and using the
$\Delta(x)$ dependence obtained in the previous step, we find the
Green functions by repeating the iterations until convergency is
reached. We define the full DOS $N(E)$ and the spin resolved DOS
$N_{\uparrow(\downarrow)}(E)$, normalized to the DOS in the normal
state, as
\begin{subequations}
\label{DOS}
\begin{align}
N(E) &= \left[ N_{\uparrow}(E) + N_{\downarrow}(E)\right]/2, \label{DOS_full} \\
N_{\uparrow(\downarrow)}(E) &= \re\left[\cos\theta_{b \uparrow(\downarrow)}(%
i\omega \rightarrow E + i0)\right],  \label{DOS_spin}
\end{align}
\end{subequations}
where $\theta_b$ is the boundary value of $\theta_f$ at $x =
-d_f/2$. In case of a long F layer ($d_f \gg \xi_{f1}$) it is also
possible to obtain an analytical expression for the DOS at the
free boundary of the ferromagnet \cite{Vasenko, Cretinon}.

In Fig.~\ref{DOSa0} and \ref{DOSa1} we plot the DOS energy
dependence for different $d_f$. Fig.~\ref{DOSa0} corresponds to
the case of the absence of spin-flip scattering (infinite
$\tau_m$) and Fig.~\ref{DOSa1} to the case of $\tau_m = 1/ \pi
T_c$. In both figures we plot the DOS for two chosen values of
exchange field, $h = \pi T_c$ for plots (a)-(d) and $h = 3\pi T_c$
for plots (e)-(h). In Fig.~\ref{DOSa0} we see that for any chosen
$h$ at small $d_f$ the full DOS turns to zero inside a mini-gap,
which vanishes with the increase of $d_f$. Then the DOS at the
Fermi energy $N(0)$ rapidly increases to the values larger than
unity and with further increase of $d_f$ it oscillates around
unity while it's absolute value exponentially approaches unity. In
the case of long enough ferromagnets we can observe DOS peaks at
$E = h$ [we notice that $\pi T_c \approx 1.79 \Delta$]. Also for
$d_f \gtrsim 1.5 \; \xi_n$ \cite{Vasenko} we can use an analytical
approximation for $N(E)$ \cite{Vasenko, Cretinon}. For smaller
$d_f$ it is incorrect to use this approximation. It is shown by
the dashed red line and is in rather good agreement with the
numerical solution. In the numerical curves the peaks at $E = h$
are smeared because we used finite $\gamma = 0.01$ for the
transparent F/S interface at $x = d_f/2$.

In Fig.~\ref{DOSa1} we observe similar tendencies. However for smaller $\tau_m$ the
mini-gap closes at smaller $d_f$, the period of the DOS oscillations at the Fermi
energy increases and the damped exponential decay occurs faster. Also the DOS peak at $E = h$
is smeared.

\begin{figure}[tb]
\epsfxsize=7.5cm\epsffile{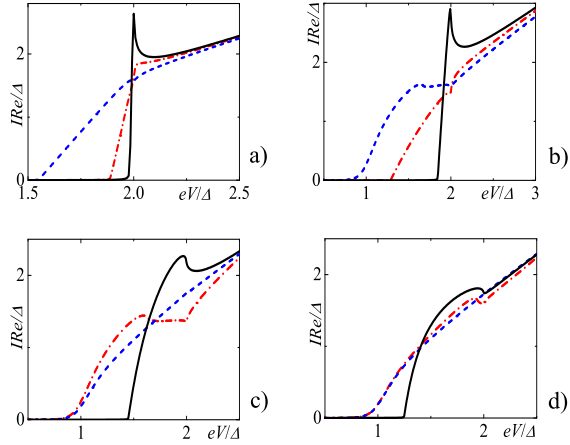} 
\caption{(Color online) Current-voltage characteristics of a SIFS junction in the absence of spin-flip scattering
for different values of the F-layer thickness $d_f$. The temperature $T = 0.1 T_c$. The exchange field
$h = 0$ (black line, which correspond to the case of a SINS junction),
$h = \pi T_c$ (red dash-dotted line), and $h = 3 \pi T_c$ (blue dashed line). (a): $d_f/\xi_n=0.5$; (b): $d_f/\xi_n=1$, (c): $d_f/\xi_n=2$,
and (d): $d_f/\xi_n=3$.}
\label{IVa0}
\end{figure}
\begin{figure}[tb]
\epsfxsize=7.5cm\epsffile{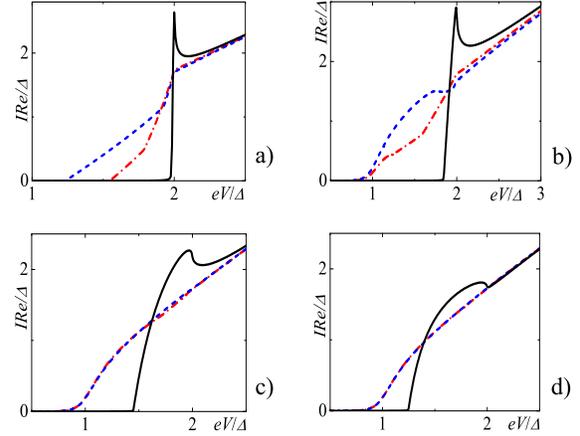} 
\caption{(Color online) Current-voltage characteristics of a SIFS junction for $\tau_m = 1/\pi T_c$. The temperature $T = 0.1 T_c$. The exchange field
$h = 0$ (black line, which correspond to the case of a SINS junction),
$h = \pi T_c$ (red dash-dotted line), and $h = 3 \pi T_c$ (blue dashed line). (a): $d_f/\xi_n=0.5$; (b): $d_f/\xi_n=1$, (c): $d_f/\xi_n=2$,
and (d): $d_f/\xi_n=3$.}
\label{IVa1}
\end{figure}
%


\section{Current-voltage characteristics of SIFS}\label{CVC_SIFS}

We calculate the current-voltage characteristics (CVC) of a SIFS junction at $T = 0.1 T_c$ using \Eq{I(V)}. Plots for different parameters are
shown in Figs.~\ref{IVa0} and \ref{IVa1}. For comparison we also present  the CVC of a SINS tunnel junction, i.e. a junction
with a normal metal interlayer instead of a ferromagnet. We can see that for a certain range of parameters the CVC of a SIFS junction
exhibits an interesting ${\cal N}$-like feature [we can observe it for $h = 3 \pi T_c$ in Fig.~\ref{IVa0} (b) and for
$h = \pi T_c$ in Fig.~\ref{IVa0} (c)]. It corresponds to the case of a large subgap DOS in Fig.~\ref{DOSa0} (c),(f). For finite spin-flip
scattering the features of CVC are smeared. At large $d_f$ they totally disappear. A more detailed analysis of CVC of SIFS junctions will be presented
elsewhere \cite{Vasenko2}.

\section{Summary}

To summarize, we calculated the quasi-particle DOS in the F layer of a SIFS junction in close
vicinity of the tunnel barrier and use it to obtain the current-voltage characteristics of a SIFS junction.
The developed formalism may be used for estimations of dissipation in SIFS-based qubits, which
will be done elsewhere.

This work was supported by Na\-no\-SciERA ``Na\-no\-fridge'' EU project and RFBR Project No. N09-02-12176.


\end{document}